\documentclass[aps,preprint,epsfig,rotate]{revtex4}
\usepackage{graphicx}
\usepackage{bm}
\usepackage{epsfig}
\begin{document}
\title{On the nuclear $(n;t)-$reaction in the three-electron 
       ${}^{6}$Li atom}

 \author{Alexei M. Frolov}
 \email[E--mail address: ]{afrolov@uwo.ca}

 \author{David M. Wardlaw}
 \email[E--mail address: ]{dwardlaw@uwo.ca}

\affiliation{Department of Chemistry\\
 University of Western Ontario, London, Ontario N6H 5B7, Canada}

\date{\today}

\begin{abstract}

The nuclear $(n;t)-$reaction of the three-electron ${}^{6}$Li atom with
thermal/slow neutrons is considered. An effective method has been developed
for determining the probabilities of formation of various atoms and ions in 
different bound states. We discuss a number of fundamental questions 
directly related to numerical computations of the final state atomic 
probabilities. A few appropriate variational expansions for atomic wave 
functions of the incident lithium atom and final helium atom and/or tritium 
negatively charged ion are discussed. It appears that the final ${}^4$He 
atom arising during the nuclear $(n,{}^{6}$Li; ${}^4$He$,t)$-reaction in the
three-electron Li atom can also be created in its triplet states. The 
formation of the quasi-stable three-electron $e^{-}_3$ during the nuclear $(n;
t)-$reaction at the Li atom is briefly discussed. Bremsstrahlung emitted by 
atomic electrons accelerated by the rapidly moving nuclear fragments from this 
reaction is analyzed. The frequency spectrum of the emitted radiation is 
investigated. 

\end{abstract}
\maketitle
\newpage

\section{Introduction}

The nuclear reaction of the ${}^{6}$Li nuclei with the thermal (and slow) 
neutrons is written in the form \cite{Kik}
\begin{eqnarray}
 {}^6{\rm Li} + n = {}^4{\rm He} + {}^3{\rm H} + 4.785 \; \; MeV \label{e1}
\end{eqnarray}
where the notations ${}^4$He and ${}^3{\rm H}$ stand for the helium nucleus 
(also the called the $\alpha-$particle, or $\alpha$, for short) and tritium (or
$t$) nucleus. The cross-section $\sigma$ of this nuclear reaction for thermal 
neutrons with $E_n \approx 0$ is very large $\sigma_{max} \approx 960 \cdot 
10^{-24}$ $cm^2$ (or 960 $barn$) \cite{WW}. The velocities of the two nuclear 
fragments formed in the reaction, Eq.(\ref{e1}), with thermal neutrons are 
$v_t \approx$ 6.03986 $a.u.$ and $v_{\alpha} \approx$ 4.52989 $a.u.$ for the 
tritium nucleus and $\alpha-$particle, respectively. In this study all 
particle velocities are given in atomic units, where $\hbar = 1, m_e = 1, e 
= 1$ and the unit of atomic velocity is $v_e = \alpha c \approx \frac{c}{137} 
\approx 2.1882661 \cdot 10^{8}$ $cm \cdot sec^{-1}$. Here and everywhere 
below $c$ is the speed of light and $\alpha = \frac{e^2}{\hbar c}$ is the 
dimensionless fine structure constant. This `atomic velocity' $v_e$ is the 
velocity of the $1s-$electron in the hydrogen atom with the infinitely heavy 
nucleus ${}^{\infty}$H. It is clear that in atomic units $v_e = 1$.

In reality, at normal conditions the nuclear reaction, Eq.(1), occurs in the 
three-electron ${}^6$Li atom. Therefore, the electron density of the original 
Li atom is re-distributed somehow between the two new nuclei created in the 
reaction, Eq.(\ref{e1}). The newly created atomic nuclei move rapidly with 
velocities which significantly exceed average velocities of atomic electrons 
in the incident Li atom. The re-distribution of electron density of the incident 
atom between two rapidly moving nuclei leads to the three following processes: (1) 
formation of the final atoms and/or ions in a variety of different excited states, 
(2) emission of breaking radiation (or bremsstrahlung) by the accelerated 
electrons, and (3) formation of the spatially isolated quasi-stable three-electron 
system $e^{-}_3$. The first goal of this study is to evaluate the final state 
(atomic) probabilities for the reaction Eq.(\ref{e1}). Briefly, we want to determine 
approximate probabilities to detect the helium and tritium atoms, and He$^{+}$ 
and ${}^{3}$H$^{-}$ ions in their bound states. Note that all newly created atomic 
fragments from the reaction, Eq.(\ref{e1}), move rapidly even in the case of thermal 
and slow neutrons. The accurate theoretical prediction of the final state 
probabilities for rapidly moving atomic fragments is not trivial. Moreover, addressing
this problem reveals a large number of unanswered fundamental questions. In this study
we have discovered, among other things, that the final ${}^4$He atom can be formed not 
only in the singlet spin states, but in a number of triplet states also. Another 
interesting possibility is the formation of the quasi-stable three-electron system 
$e^{-}_3$ during the nuclear reaction, Eq.(\ref{e1}). The $e^{-}_3$ system contains no 
positive (heavy) particles and has a number of unique properties. In this study we 
also want to analyze the electromagnetic radiation (or bremsstrahlung) emitted after 
the nuclear reaction, Eq.(\ref{e1}), in the three-electron Li atom. It is shown below 
that the frequency spectrum of such radiation essentially coincides with the well 
known bremsstrahlung spectrum.  

\section{Sudden approximation and electron density matrices}

As we have mentioned above the rate of the nuclear reaction, Eq.(\ref{e1}), is 
substantially faster (10,000 times faster) than the rates of atomic transitions in the 
incident Li atom. Moreover, the two new nuclei (${}^4$He and ${}^3$H) arising in this 
reaction move with the velocities which are significantly larger than the corresponding 
electron velocities in the incident Li atom. This means that to determine the final 
state probabilities (i.e. probabilities to find newly formed atomic systems in some 
final states) we can apply the so-called sudden approximation \cite{Mig1}, \cite{LLQ}, 
\cite{FrWa2007}. Note also that after the nuclear reaction, Eq.(\ref{e1}), in the 
three-electron Li atom we have to deal with three new sub-systems. Two of these systems 
are the positively charged ${}^4$He and ${}^3$H nuclei rapidly moving from the reaction 
area (fast sub-systems). The third (slow) sub-system includes three electrons remaining 
in the reaction area, i.e. the three-electron $e^{-}_3$ (or tri-electron). These three 
electrons are left from the incident Li atom and their wave functions still keep 
information about the original atom which disappeared during the nuclear reaction, 
Eq.(\ref{e1}). Interactions between these three sub-systems produce all phenomena 
mentioned above, i.e. formation of the different atomic species in a variety of bound 
(and unbound) states and emission of breaking radiation.  

Suppose we want to determine the probability that the rapidly moving ${}^{4}$He 
nucleus takes one of the three atomic (Li) electrons remaining unbound after the nuclear 
reaction, Eq.(\ref{e1}). In other words, the final atomic system is the positively 
charged helium ion ${}^4$He$^{+}$. For an atomic system containing fast and slow 
sub-systems the corresponding probability is written in the form (see, e.g., 
\cite{FrWa2007})
\begin{eqnarray}
 P_{if} = \mid \int \int \rho^{(1)}_{\rm Li}({\bf r}_{14}, {\bf r}^{\prime}_{14})
 \exp(\imath {\bf V}_{\alpha} \cdot {\bf r}_{14} - \imath {\bf V}_{\alpha} \cdot 
 {\bf r}^{\prime}_{14})
 \rho_{\rm He^{+}}({\bf r}_{14}, {\bf r}^{\prime}_{14}) d^{3}{\bf r}_{14} 
 d^{3} {\bf r}^{\prime}_{14} \mid \label{intg1}
\end{eqnarray}
where ${\bf V}_{\alpha}$ is the velocity of the He$^{+}$ ion after the reaction, 
Eq.(\ref{e1}) and $\rho^{(1)}_{\rm Li}({\bf r}_{14}, {\bf r}^{\prime}_{14})$ and 
$\rho_{\rm He^{+}}({\bf r}_{14}, {\bf r}^{\prime}_{14})$ are the generalized 
one-electron density matrixes of the Li atom and He$^{+}$ ion, respectively. The exponential 
factor in this formula is the appropriate Galilean factor. The expression, Eq.(\ref{intg1}), 
follows from the formula for the Galilean transformation ${\bf r} \rightarrow {\bf r} - 
{\bf V} t$ of the wave function of a single particle with the mass $m$
\begin{eqnarray}
 \Psi({\bf r}, t) = \Psi^{\prime}({\bf r} - {\bf V} t, t) exp\Bigl[ \frac{\imath m}{\hbar}
 (-{\bf V} \cdot {\bf r} - \frac12 V^2 t) \Bigr]
\end{eqnarray}
where $V = \mid {\bf V} \mid$. In the limt $t \rightarrow 0$ and in atomic units $\hbar = 1, 
m_e = 1, e = 1$ one finds from this formula for a one-electron wave function
\begin{eqnarray}
 \Psi^{\prime}({\bf r}) = \Psi({\bf r}) exp(\imath {\bf V} \cdot {\bf r})
\end{eqnarray}
where $\Psi^{\prime}({\bf r})$ is the spatial wave function of the system moving with the
constant speed ${\bf V}$. The formula, Eq.(\ref{intg1}), for three-electron wave function 
follows from the last equation. 

The generalized one-electron density matrices in Eq.(\ref{intg1}) are simply related with the 
corresponding wave functions of the bound states, e.g., for the Li atom
\begin{eqnarray}
 \rho^{(1)}_{\rm Li}({\bf r}_{14}, {\bf r}^{\prime}_{14}) = 3 \int \int \int \int
 \Psi^{*}_{\rm Li}({\bf x}_{14}, {\bf x}_{24}, {\bf x}_{34}) \Psi_{\rm Li}({\bf 
 x}^{\prime}_{14}, {\bf x}_{24}, {\bf x}_{34}) ds_{1} ds_{2} d{\bf x}_{24} 
 d{\bf x}_{34} \label{intg2}
\end{eqnarray}
where ${\bf x}_i = (s_i, {\bf r}_i)$ is the complete set of spin-spatial coordinates of the 
$i$-th electron. In Eqs.(\ref{intg1}) - (\ref{intg2}) and everywhere below the nucleus in the 
three-electron Li atom is designated by the index 4, while the three electrons are denoted by 
the indexes 1, 2 and 3. All few-electron wave functions in this equation and below are 
assumed to be properly symmetrized in respect to all electron coordinates. This definition of 
the generalized one-electron density matrix, Eq.(\ref{intg2})), and notations used in this 
equation correspond to the definition given in Eqs.(4.1.5) - (4.1.6) of \cite{MQW}. The 
generalized one-electron density matrix for the one-electron He$^{+}$ ion is 
\begin{eqnarray}
 \rho_{\rm He^{+}}({\bf r}_{14}, {\bf r}^{\prime}_{14}) 
 = \rho^{(1)}_{\rm He^{+}}({\bf r}_{14}, {\bf r}^{\prime}_{14}) 
 = \int \int \Phi^{*}_{\rm He^{+}}({\bf x}_{14}) 
 \Phi_{\rm He^{+}}({\bf x}^{\prime}_{14}) ds_{1} \label{intg3}
\end{eqnarray}
It contains only one integral over spin variables of the electron 1. 

If the tritium atom is formed during the nuclear reaction, Eq.(\ref{e1}) in the three-electron 
Li atom, then we can write the following formula for the final state probability $P_{if}$
\begin{eqnarray}
 P_{if} = \mid \int \int \rho^{(1)}_{\rm Li}({\bf r}_{14}, {\bf r}^{\prime}_{14})
 \exp(\imath {\bf V}_{t} \cdot {\bf r}_{14} - \imath {\bf V}_{t} \cdot 
 {\bf r}^{\prime}_{14}) \rho_{\rm T}({\bf r}_{14}, {\bf r}^{\prime}_{14}) d^{3}{\bf r}_{14} 
 d^3{\bf r}^{\prime}_{14} \mid \label{intg4}
\end{eqnarray}
where $\rho_{\rm T}({\bf r}_{14}, {\bf r}^{\prime}_{14})$ is the generalized one-electron 
density matrix of the tritium atom. and ${\bf V}_{t}$ is the velocity of the tritium nucleus 
after the nuclear reaction, Eq.(\ref{e1}). The definition of the $\rho_{\rm T}({\bf r}_{14}, 
{\bf r}^{\prime}_{14})$ matrix corresponds to the definition given in Eq.(\ref{intg3}). In 
Eq.(\ref{intg3}) the notations T and $t$ stand for the tritium and and tritium nucleus, 
respectively, while ${\bf V}_{t}$ is the velocity of the tritium atom and tritium nucleus 
after the nuclear reaction, Eq.(\ref{e1}).

For the helium atom product (two bound electrons) the final state probability $P_{if}$ is 
written in the form
\begin{eqnarray}
 P_{if} = \int \int \int \int \rho^{(2)}_{\rm Li}({\bf r}_{14}, {\bf r}_{24}; 
 {\bf r}^{\prime}_{14}, {\bf r}^{\prime}_{24}) \exp[\imath {\bf V}_{\alpha} \cdot ({\bf r}_{14} - 
 {\bf r}^{\prime}_{14}) + \imath {\bf V}_{\alpha} \cdot ({\bf r}_{24} - 
 {\bf r}^{\prime}_{24})] \nonumber \\
 \rho^{(2)}_{\rm He}({\bf r}_{14}, {\bf r}_{24}; 
 {\bf r}^{\prime}_{14}, {\bf r}^{\prime}_{24}) d^{3}{\bf r}_{14} d^{3}{\bf r}_{24} 
 d^{3}{\bf r}^{\prime}_{14} d^{3}{\bf r}^{\prime}_{24} \label{intg5}
\end{eqnarray}
where ${\bf V}_{\alpha}$ is the final velocity of the $\alpha-$particle after the nuclear
reaction, Eq.(\ref{e1}). The generalized two-electron density matrix of the three-electron Li 
atom is defined as follows 
\begin{eqnarray}
 \rho^{(2)}_{\rm Li}({\bf r}_{14}, {\bf r}_{24}; {\bf r}^{\prime}_{14}, {\bf r}^{\prime}_{24}) 
 = 6 \int \int \int \int \int \Psi^{*}_{\rm Li}({\bf x}_{14}, {\bf x}_{24}, {\bf x}_{34}) 
 \Psi_{\rm Li}({\bf x}^{\prime}_{14}, {\bf x}^{\prime}_{24}, {\bf x}_{34}) 
 ds_{1} ds_{2} ds^{\prime}_{1} ds^{\prime}_{2} d{\bf x}_{34} \label{intg6}
\end{eqnarray}
where we have used the factor $N (N - 1)$ in the front of the generalized density matrices 
\cite{MQW}. In some cases, however, it is better to use a different definition of the 
density matrices with the unity factor in front of them. The generalized two-electron density 
matrix of the He atom (bound states) takes the form
\begin{eqnarray}
 \rho^{(2)}_{\rm He}({\bf r}_{14}, {\bf r}_{24}; {\bf r}^{\prime}_{14}, {\bf r}^{\prime}_{24}) 
 = 2 \int \int \int \int \Phi^{*}_{\rm He}({\bf x}_{14}, {\bf x}_{24}) 
 \Phi_{\rm He}({\bf x}^{\prime}_{14}, {\bf x}^{\prime}_{24}) ds_{1} ds_{2} ds^{\prime}_{1} 
 ds^{\prime}_{2} \label{intg7}
\end{eqnarray}

In the case of nuclear reaction, Eq.(\ref{e1}), in the three electron Li atom we need only
one- and two-electron generalized density matrices, since the negatively charged helium ion 
(He$^{-}$) is unstable. Note that all these integrals mentioned in the expressions for the 
probabilities, Eq.(\ref{intg1}), Eqs.(\ref{intg4}) - (\ref{intg5}) and generalized density 
matrices, Eqs.(\ref{intg2}) - (\ref{intg3}) and (\ref{intg7}), include the wave functions of 
the bound few-electron atoms and ions. These wave functions must be determined from accurate 
atomic computations, whose construction is discussed in the fourth Section.

\section{Alternative approach based on the wave functions}

The approach described above is based on the explicit construction of the few-electron 
density matrices. This method is useful for general theoretical analysis. However, in 
numerical applications, e.g., to determine the final state probabilities of the actual 
processes, it becomes a very complicated procedure. In general, even approximate evaluations 
with this method require an extensive analytical and computational work. In many cases, the 
method based on the use of the density matrices leads to a substantial loss of the numerical 
accuracy. Therefore, it is crucially important to develop another (or alternative) method 
which can be used for accurate numerical evaluations of the final state probabilities. After 
extensive numerical research we developed a new method for approximate evaluation of 
the final state probabilities in the case of the reaction, Eq.(\ref{e1}), and for other 
similar processes. This method is based on the use of the wave functions (i.e. it does not 
need any density matrix). Our procedure and its numerous advantages are described here. 

In the previous Section we have assumed that all wave functions used in this study are the 
truly correlated wave functions obtained from highly accurate computations of the 
corresponding few-electron systems (Li atom, He atom, etc). Such wave functions are easily 
obtained in modern numerical computations of the bound state spectra of few-electron atomic 
systems. However, if some part of the incident system begins to move after the nuclear 
reaction (see, e.g., Eq.(\ref{e1})), then we need to determine the partial (or complete) 
Fourier transformation of the incident wave function. In general, for the truly correlated 
wave functions this is a very difficult proposition, since such a wave function includes not 
only electron-nuclear coordinates, but also all electron-electron coordinates (or correlation 
coordinates). It is clear that this problem will be avoided if we can use the wave functions 
which depend upon the electron-nuclear coordinates only. Indeed, let $\Psi_{Li}(r_{1}, r_{2}, 
r_{3}) (= \Psi(r_{14}, r_{24}, r_{34}))$ be such a variational wave function of the ground 
${}^2S-$state of the Li atom. This wave function is represented in the following form
\begin{eqnarray}
 \Psi_{Li}(r_1, r_2, r_3) = \sum^{K}_{k=1} \sum^{M}_{m=1} \sum^{N}_{n=1} C_{kmn} \phi_k(r_1)
 \phi_m(r_2) \phi_n(r_3) \label{psi1}
\end{eqnarray}
where $C_{kmn}$ are the numerical coefficients, while $\phi_{i}(r)$ are the unit-norm radial 
basis functions which form a complete set of pair-wise orthogonal functions defined on the 
semi-interval $[0, +\infty)$. Let us assume that the wave function of the final He atom from 
the reaction Eq.(\ref{e1}) is represented in the similar form
\begin{eqnarray}
 \Psi_{He}(r_1, r_2) = \sum^{K}_{k=1} \sum^{M}_{m=1} B_{km} \psi_k(r_1) \psi_m(r_2) 
 \label{psi2}
\end{eqnarray}
All these wave functions $\Psi_{Li}(r_1, r_2, r_3)$ and $\Psi_{He}(r_1, r_2)$ must be
properly symmetrized upon all electron variables. Furthermore, without loss of generality we 
shall assume that these wave functions have unit norms.  

As we mentioned above the final He atom formed in the reaction Eq.(\ref{e1}) is moving with 
the constant speed $V$. Our first goal in this Section is to evaluate the probability to form 
the two-electron He atom during the reaction, Eq.(\ref{e1}). With the factorized wave functions, 
Eqs.(\ref{psi1}) and (\ref{psi2}), the final state probability can be determined as the result 
of the following two-stage procedure. At the first stage by calculating the integral
\begin{eqnarray}
 \int_0^{+\infty} \Psi_{Li}(r_1, r_2, r_3) \phi_n(r_3) r^2_3 dr_3 = \overline{\Psi}_{Li}(r_1, 
 r_2) \label{psi3}
\end{eqnarray}
we determine the two-electron sub-wave function $\overline{\Psi}_{Li}(r_1, r_2)$ of the Li 
atom. At the second stage of the method we need to compute the following integral
\begin{eqnarray}
 M = \int_{0}^{+\infty} \int_{0}^{+\infty} \Psi_{He}(r_1, r_2) j_0(V_{He} r_1) j_0(V_{He} r_2)
 \overline{\Psi}_{Li}(r_1, r_2) r^2_1 dr_1 r^2_2 dr_2 \label{psi4}
\end{eqnarray}
which coincides with the probability amplitude. The actual computations are reduced to the
numerical summation of some special one-dimensional integrals with the spherical Bessel functions
$j_0(x)$. In general, the probability amplitude $M$, Eq.(\ref{psi4}), may include other Bessel 
functions $j_L(V r_i)$ with $L \ge 1$, depending upon the final state of the He atom, but we 
do not want to discuss this aspect here. In reality, all such calculations are substantially 
simpler than computations based on the use of the density matrices (see Section II above). The 
advantage of the factorized wave functions, Eqs.(\ref{psi3}) - (\ref{psi4}), for our problems is 
obvious. Indeed, such functions approximate the bound state wave functions to a sufficiently high 
accuracy. On the other hand, for these functions we can easily perform partial and/or complete 
Fourier transformations which are needed to determine the final state probabilities.   

For the He$^{+}$ ion product in the reaction Eq.(\ref{e1}) the expression for the probability
amplitude is written in the form 
\begin{eqnarray}
 M = \int_{0}^{+\infty} \Psi_{He^{+}}(r_1) j_0(V r_1) \overline{\Psi}_{Li}(r_1) r^2_1 dr_1 
 \label{psi5}
\end{eqnarray}
where the one-electron sub-wave function $\overline{\Psi}_{Li}(r_1)$ of the Li atom is 
\begin{eqnarray}
 \overline{\Psi}_{Li}(r_1) = \int_0^{+\infty} \int_0^{+\infty} \Psi_{Li}(r_1, r_2, r_3) 
 \phi_m(r_2) \phi_n(r_3) r^2_2 dr_2 r^2_3 dr_3 \label{psi6}
\end{eqnarray}
For the tritium atom T product the probability amplitude takes analogous form 
\begin{eqnarray}
 M = \int_{0}^{+\infty} \Psi_{T}(r_1) j_0(V r_1) \overline{\Psi}_{Li}(r_1) r^2_1 dr_1 
 \label{psi7}
\end{eqnarray}
Note again that all analytical and numerical computations with the factorized wave functions
are simple and straightforward. Our method of construction of the approximate few-electron
wave functions corresponds to the model of independent (or quasi-independent) electrons.
Nevertheless, it provides sufficient numerical accuracy in actual applications, e.g., to the
nuclear reaction, Eq.(\ref{e1}) (see below).

\section{Variational wave function of the lithium atom}

To determine the final state probabilities we need to construct the accurate 
variational wave functions for the incident and final atomic systems involved in the 
reaction, Eq.(\ref{e1}). In our earlier paper \cite{FrWa09} we have calculated the final 
state probabilities to form one-electron atoms and ions in those cases when exothermic 
nuclear $(n;t)-$ and $(n;\alpha)-$reactions occur in one-electron atoms/ions, i.e., both 
incident and final atomic systems contain only one bound electron. In this study we deal 
with the actual three-electron wave function of the incident ${}^6$Li atom and two-electron 
wave function of the final ${}^4$He atom. Also, we consider the formation of the 
one-electron He$^{+}$ ion, tritium atom ${}^3$H (or T) and two-electron negatively charged 
tritium ion ${}^3$H$^{-}$ (or T$^{-}$).

First, let us discuss the construction of the three-electron variational wave function
for the Li atom. Without loss of generality, below we restrict ourselves to the 
consideration of the ground ${}^2S(L = 0)-$state of the Li atom. As is well known (see, e.g., 
\cite{Lars}, \cite{Fro2011}) the accurate variational wave function of the ground (doublet) 
${}^2S(L = 0)-$state of the Li atom is written in the following general form
\begin{eqnarray}
 \Psi({\rm Li})_{L=0} = \psi_{L=0}(A; \bigl\{ r_{ij} \bigr\}) (\alpha \beta
 \alpha - \beta \alpha \alpha) + \phi_{L=0}(B; \bigl\{ r_{ij} \bigr\}) (2
 \alpha \alpha \beta  - \beta \alpha \alpha - \alpha \beta \alpha)
 \label{psi}
\end{eqnarray}
where $\psi_{L=0}(A; \bigl\{ r_{ij} \bigr\})$ and $\phi_{L=0}(B; \bigl\{
r_{ij} \bigr\})$ are the two independent radial parts (= spatial parts) of
the total wave function. Everywhere below in this study, we shall assume
that all mentioned wave functions have unit norms. The notations $\alpha$ and
$\beta$ in Eq.(\ref{psi}) stand for the one-electron spin-up and spin-down
functions, respectively (see, e.g., \cite{Dir}). The notations $A$ and $B$
in Eq.(\ref{psi}) mean that the two sets of non-linear parameters associated
with the radial functions $\psi$ and $\phi$ can be optimized independently. Note
that each of the radial basis functions in Eq.(\ref{psi}) explicitly depends
upon all six interparticle (or relative) coordinates $r_{12}, r_{13},
r_{23}, r_{14}, r_{24}, r_{34}$, where the indexes 1, 2, 3 stand for the 
three electrons, while index 4 means the nucleus.
 
The actual atomic wave function in an atomic system must be completely 
antisymmetric with respect to all electron spin-spatial variables. For a
three-electron wave function this requirement is written in the form
${\hat{\cal A}}_{123} \Psi(1,2,3) = - \Psi(1,2,3)$, where the wave function 
$\Psi$ is given by Eq.(\ref{psi}) and $\hat{{\cal A}}_e$ is the three-particle 
(or three-electron) antisymmetrizer ${\hat{\cal A}}_e = \hat{e} - \hat{P}_{12} 
- \hat{P}_{13} - \hat{P}_{23} + \hat{P}_{123} + \hat{P}_{132}$. Here $\hat{e}$ 
is the identity permutation, while $\hat{P}_{ij}$ is the permutation of the 
$i$-th and $j$-th particles. Analogously, the operator $\hat{P}_{ijk}$ is the
permutation of the $i$-th, $j$-th and $k$-th particles. 

Suppose that the three-electron wave function of the Li atom has been constructed in 
the form of Eq.(\ref{psi}). By applying the antisymmetrizer ${\hat{\cal A}}_{123}$ to 
the first part of the total wave function, Eq.(\ref{psi}), one finds
\begin{eqnarray}
 {\hat{\cal A}}_{123} \Bigl[ \psi_{L=0}(A; \bigl\{ r_{ij} \bigr\}) (\alpha \beta
 \alpha - \beta \alpha \alpha) \bigr] = (\hat{e} \psi) (\alpha \beta
 \alpha - \beta \alpha \alpha) 
 + (\hat{P}_{12} \psi) (\alpha \beta \alpha - \beta \alpha \alpha) \nonumber \\
 - (\hat{P}_{13} \psi) (\alpha \beta \alpha - \alpha \alpha \beta)
 - (\hat{P}_{23} \psi) (\alpha \alpha \beta - \beta \alpha \alpha) 
 + (\hat{P}_{123} \psi) (\alpha \alpha \beta - \alpha \beta \alpha)\nonumber \\
 + (\hat{P}_{132} \psi) (\beta \alpha \alpha - \alpha \alpha \beta)
 \label{psi1}
\end{eqnarray}
where the notations $(\hat{P}_{ij} \psi)$ and $(\hat{P}_{ijk} \psi)$ mean the 
permutation operators which act on the coordinate wave functions only. Analogously, 
for the second part of the total wave function one finds
\begin{eqnarray}
 {\hat{\cal A}}_{123} \Bigl[ \phi_{L=0}(B; \bigl\{ r_{ij} \bigr\}) (2 \alpha 
 \alpha \beta - \beta \alpha \alpha - \alpha \beta \alpha) \bigr] = 
 (\hat{e} \phi) (2 \alpha \alpha \beta - \beta \alpha \alpha - \alpha \beta \alpha) 
 \nonumber  \\
 - (\hat{P}_{12} \phi) (2 \alpha \alpha \beta - \beta \alpha \alpha - 
 \alpha \beta \alpha)
 - (\hat{P}_{13} \phi) (2 \beta \alpha \alpha - \alpha \alpha \beta -
 \alpha \beta \alpha) \nonumber \\
 - (\hat{P}_{23} \phi) (2 \alpha \beta \alpha - \alpha \alpha \beta -
  \beta \alpha \alpha)
 + (\hat{P}_{123} \phi) (2 \beta \alpha \alpha - \alpha \beta \alpha -
  \alpha \alpha \beta) \nonumber \\
 + (\hat{P}_{132} \phi) (2 \alpha \beta \alpha - \alpha \alpha \beta -
  \beta \alpha \alpha) \label{phi1}
\end{eqnarray}
where the notations $(\hat{P}_{ij} \phi)$ and $(\hat{P}_{ijk} \phi)$ mean the 
permutations of the spatial coordinates in the $\phi_{L=0}(B; \bigl\{ r_{ij} 
\bigr\})$ radial function, Eq.(\ref{psi}). 

Now, by using the formulas, Eqs.(\ref{psi1}) and (\ref{phi1}), we can obtain the
formulas which can be used in computations of the final state probabilities in the
case of nuclear reaction, Eq.(\ref{e1}), in the three-electron Li atom. For 
instance, if the final wave function has the same spin-symmetry, i.e. it is written
in the form
\begin{eqnarray}
 \Psi_{fi} = \psi_{fi}({\bf r}_{1}, {\bf r}_{2}, {\bf r}_3) (\alpha \beta
 \alpha - \beta \alpha \alpha) + \phi_{fi}({\bf r}_{1}, {\bf r}_{2}, {\bf r}_3) 
 (2 \alpha \alpha \beta  - \beta \alpha \alpha - \alpha \beta \alpha) \label{psif}
\end{eqnarray}
then the final state probabilities are determined with the use of the following 
formulas 
\begin{eqnarray}
 P_{\psi\psi} = \langle \psi_{fi}({\bf r}_{1}, {\bf r}_{2}, {\bf r}_3) \mid 
 \frac{1}{2 \sqrt{3}} \Bigl( 2 \hat{e} + 2 \hat{P}_{12} - \hat{P}_{13} - \hat{P}_{23} 
 - \hat{P}_{123} - \hat{P}_{132} \Bigr) \psi_{{\rm Li}}(A; \bigl\{ r_{ij} \bigr\}) 
 \rangle \label{spin1} \\
 P_{\phi\psi} = \langle \phi_{fi}({\bf r}_{1}, {\bf r}_{2}, {\bf r}_3) \mid 
 \frac12 \Bigl( \hat{P}_{13} - \hat{P}_{23} + \hat{P}_{123} - \hat{P}_{132} \Bigr) 
 \psi_{{\rm Li}}(A; \bigl\{ r_{ij} \bigr\}) 
 \rangle \label{spin2} \\
 P_{\psi\phi} = \langle \psi_{fi}({\bf r}_{1}, {\bf r}_{2}, {\bf r}_3) \mid 
 \frac12 \Bigl( \hat{P}_{13} - \hat{P}_{23} + \hat{P}_{123} - \hat{P}_{132} \Bigr) 
 \phi_{{\rm Li}}(B; \bigl\{ r_{ij} \bigr\}) 
 \rangle \label{spin3} \\
 P_{\phi\phi} = \langle \phi_{fi}({\bf r}_{1}, {\bf r}_{2}, {\bf r}_3) \mid 
 \frac{1}{2 \sqrt{3}} \Bigl( 2 \hat{e} - 2 \hat{P}_{12} + \hat{P}_{13} + \hat{P}_{23} 
 - \hat{P}_{123} - \hat{P}_{132} \Bigr) \phi_{{\rm Li}}(B; \bigl\{ r_{ij} \bigr\})
 \rangle \label{spin4}
\end{eqnarray}
Note that these formulas coincide with the known formulas \cite{FrWa2010} which 
correspond to the doublet $\rightarrow$ doublet transition in the three-electron
atomic systems. Briefly, this means that the both incident and final atomic states
contain three-electrons in the doublet spin configuration (the total electron spin
equals $\frac12$). 

By considering all possible spin configurations for the final atomic state one finds
that other spin configurations are possible. For instance, if reaction Eq.(\ref{e1}) 
leads to the formation of the two neutral atoms (${}^4$He and ${}^3$H atoms), then 
the helium atom can be detected either in the singlet state, or in the triplet state. 
The spin function of the singlet state is $\chi_s = \frac{1}{\sqrt{2}} (\alpha \beta 
- \beta \alpha)$, while the wave function of the triplet state can be chosen (in our 
case) in the form $\chi_t = \alpha \alpha$. It is clear that these two wave functions 
have unit norms and they are orthogonal to each other. The probability of formation 
of the ${}^4$He atom in its singlet spin state is determined by the fourmulas, 
Eqs.(\ref{spin1}) - (\ref{spin4}). The tritium atom in this case will be formed with 
the $\beta-$electron spin function.

If the final state of the ${}^4$He atom is the triplet state ($tr$), then the formulas 
for the final state probabilities take the form
\begin{eqnarray}
 P_{tr;\psi} = \langle \psi_{fi}({\bf r}_{1}, {\bf r}_{2}, {\bf r}_3) \mid 
 \frac{1}{2} \Bigl(\hat{P}_{13} - \hat{P}_{23} + \hat{P}_{123} - \hat{P}_{132} 
 \Bigr) \psi_{{\rm Li}}(A; \bigl\{ r_{ij} \bigr\}) 
 \rangle \label{spinx} \\
 P_{tr;\phi} = \langle \psi_{fi}({\bf r}_{1}, {\bf r}_{2}, {\bf r}_3) \mid 
 \frac{1}{2 \sqrt{3}} \Bigl( 2 \hat{e} - 2 \hat{P}_{12} + \hat{P}_{13} + \hat{P}_{23} 
 - \hat{P}_{123} - \hat{P}_{132} \Bigr) \phi_{{\rm Li}}(B; \bigl\{ r_{ij} \bigr\}) 
 \rangle \label{spiny} 
\end{eqnarray}
These formulas indicate clearly that the probability to find the ${}^4$He atom arising
during the nuclear reaction, Eq.(\ref{e1}), in its triplet spin state(s) is not zero. In 
all earlier studies the transitions to the final atomic states with different spin states 
was ignored (never considered).

\section{Formation of the tri-electron}

As mentioned above the two nuclear fragments formed in the reaction, Eq.(\ref{e1}),
move with very large velocities. They leave the reaction zone very quickly for a time
which is significantly shorter than a typical atomic time $\tau_a = \frac{\hbar^3}{m_e e^4}
\approx 2.42 \cdot 10^{-17}$ $sec$. The three negatively charged electrons $e^{-}$ remain 
in the reaction zone substantially longer. The arising quasi-stable system of three 
electrons $e^{-}_3$ is called a three-electron, or tri-electron. It is clear that the 
tri-electron is not a stable system and after some time it will be destroyed by the 
Coulomb repulsion between electrons. However, in some cases such a destruction takes a 
relatively long time which is sufficient to detect it experimentally and measure 
some of its properties. A combination of the stabilizing electric and magnetic fields can 
be used to increase the life-time of a tri-electron. Some of the properties of the $e^{-}_3$ 
system can be studied by using modern methods of the femto- and attosecond physics. 
Experimental interest to study such new systems cannot be overestimated.

The `quasi-atomic' tri-electron system has a large number of unique properties and 
can be considered as a light quasi-atom which has no central (i.e. heavy), positively 
charged nucleus. In the case of reaction, Eq.(\ref{e1}), the spin-spatial symmetry of the 
three-electron wave function exactly coincides with the corresponding symmetry of the wave 
function of the ground ${}^{2}S$-state of the Li atom. 

\section{Calculations}

In this study we have developed an approximate procedure to construct the bound state wave
functions of the three-electron atoms and ions. This procedure can be used to perform 
numerical evaluations of the final state probabilities in the case of the nuclear reaction, 
Eq.(\ref{e1}), and for the $\beta^{-}$-decaying isotopes of the three-electron atoms 
\cite{FrBe}. In this approach the trial wave function is constructed as the sum of various 
terms and each of these terms contains the products of the electron-nucleus functions. None 
of the three electron-electron coordinates $r_{32}, r_{31}, r_{21}$ is included in such 
trial wave functions. This simplifies drastically the following Fourier transforms of such
wave functions. For the ground state (the doublet ${}^2S(L = 0)-$state) of the Li atom the 
radial wave function $\psi_{L=0}(A; \bigl\{ r_{ij} \bigr\})$ is chosen in the following 
form:
\begin{eqnarray}
 \psi_{L=0}(r_{14}, r_{24}, r_{34}, 0, 0, 0) = \sum^{N_s}_{k=1}
 C_k r^{m_1(k)}_{14} r^{m_2(k)}_{24} r^{m_3(k)}_{34}
 exp(-\alpha_{k} r_{14} -\beta_{k} r_{24} -\gamma_{k} r_{34})
 \label{exp1} \\
 = \sum^{N_s}_{k=1} C_k r^{m_1(k)}_{1} r^{m_2(k)}_{2} r^{m_3(k)}_{3}
 exp(-\alpha_{k} r_{1} -\beta_{k} r_{2} -\gamma_{k} r_{3}) \nonumber
\end{eqnarray}
where $C_k$ are the linear (or variational) coefficients, while $m_1(k),m_2(k)$ and $m_3(k)$ 
are the three integer (non-negative) parameters, which are, in fact, the powers of the three 
electron-nucleus coordinates $r_{i4} = r_i$ ($i$ = 1, 2, 3). The real (and non-negative) 
parameters $\alpha_{k}, \beta_{k}, \gamma_{k}$ are the $3 N_s$ varied parameters of the 
variational expansion, Eq.(\ref{exp1}). Below, we shall assume that the trial wave function
Eq.(\ref{exp1}) has a unit norm. Furthermore, in all calculations performed for this study 
only one spin function $\chi_1 (\chi_1 = \alpha \beta \alpha - \beta \alpha \alpha$) is used.  
The wave function, Eq.(\ref{exp1}), must be properly anti-symmetrized upon all spin-spatial
coordinates of the three electrons. 

The principal question for the wave function, Eq.(\ref{exp1}), is related to its overall 
accuracy. If (and only if) such accuracy is relatively high, then such a wave function, 
Eq.(\ref{exp1}), can be used in actual computations of the probability amplitudes. In this 
study we have constructed the 23-term variational wave function shown in Table I. This wave 
function is represented in the form of Eq.(\ref{exp1}) and contains no electron-electron
coordinates. All sixty nine (69 = 3 $\times$ 23) non-linear parameters $\alpha_{k}, \beta_{k}, 
\gamma_{k}$ ($k$ = 1, 2 $\ldots$ 23) in this wave function have been optimized carefully in 
a series of bound state computations performed for the ground state of the Li atom. Finally, 
the total energy $E$ of the ground ${}^2S-$state of the ${}^{\infty}$Li atom obtained with 
this 23-term trial wave function, Eq.(\ref{exp1}), was -7.44859276608 $a.u.$ This energy
is close to the exact total energy of the ground state of the ${}^{\infty}$Li atom. It 
indicates a good overall quality of our approximate wave function with 23 terms which 
does not include any of the electron-electron coordinates $r_{12}, r_{13}, r_{23}$. 
This wave function is used in the computations of the final state probabilities (see 
below) for the nuclear reaction, Eq.(\ref{e1}), in three-electron Li atom. 

Note also that in atomic physics based on the Hartree-Fock and hydrogenic approximations 
the ground state in the Li atom is designated as the $2^2S-$state, while in the 
classification scheme developed in highly  accurate computations the same state is 
designated as the $1^2S-$state because it corresponds to the first eigenvalue of the 
Hamiltonian matrix $\hat{H}$. This classification scheme is very convenient for truly
correlated few-electron wave functions which represent situation where there are no
good hydrogenic quantum numbers. Nevertheless, to avoid conflicts between these two 
classification schemes in this study we follow the system of notation used earlier by 
Larsson \cite{Lars} which designated this state of the Li atom as the `ground ${}^2S$-state'. 

The advantage of the basis functions, Eq.(\ref{exp1}), for numerical calculations of 
the final state probabilities for the reaction Eq.(\ref{e1}) is two-fold: (1) these trial 
wave functions allow one to obtain relatively accurate approximations for atomic 
three-electron wave functions; and (2) all required Fourier transforms are easily performed 
for such trial wave functions since they do not include electron-electron coordinates. Note 
also, that approximate two-electron wave functions of the He atom and/or H$^{-}$ ion can be 
constructed in analogous form which contain only electron-nucleus coordinates, e.g., 
\begin{eqnarray}
 \psi_{L=0}(r_{1}, r_{2}, 0) = \frac{1}{\sqrt{2}} (1 + \hat{P}_{12}) \sum^{N_s}_{k=1}
 C_k r^{m_1(k)}_{1} r^{m_2(k)}_{2} exp(-\alpha_{k} r_{1} -\beta_{k} r_{2})
 \label{exp2} 
\end{eqnarray}

Some probabilities (in \%) of observing selected electron final states in the helium ion 
${}^4$He$^{+}$ and tritium atom ${}^3$H arising in the exothermic nuclear reaction 
Eq.(\ref{e1}) of the three-electron lithium-6 atom can be found in Table II. In 
calculations included in Table II we have assumed that the incident Li atom was in its 
ground ${}^2S-$state. Also, we assume that the nuclear $(n;t)-$reaction, Eq.(\ref{e1}), 
proceeds is produced by thermal and/or slow neutrons.    

\section{Bremsstrahlung}

As we have mentioned above the two positively charged atomic nuclei created in the 
nuclear reaction, Eq.(\ref{e1}), with thermal neutrons move with substantial velocities 
in the two opposite directions. In general, these two positively charged nuclei will 
produce acceleration of electrons which still remain in the reaction zone. In turn, the 
accelerated electrons will generate emission of radiation. It appears that the spectrum 
of this radiation essentially coincides with the known spectrum of bremsstrahlung (or 
breaking radiation). In this Section we discuss the basic properties of radiation 
emitted during the nuclear reaction, Eq.(\ref{e1}), in three-electron ${}^6$Li atom. 

Note that the reaction, Eq.(\ref{e1}), is a fission-type reaction in a
three-electron atomic system. The original nucleus decays into two electrically
charged fragments which are rapidly moving away from each other. The general 
theory of radiation emitted during such processes was developed in our earlier 
studies (see, e.g., \cite{FrWa09}, \cite{Fro07} and references therein).  
Atomic electrons from outer electronic shells become free during this 
fission-type nuclear reaction in an atomic nucleus. These free electrons interact 
with the rapidly moving nuclear fragments and such an interaction can produce 
breaking radiation or bremsstrahlung. Here we 
want to derive the explicit formulas for this radiation and investigate its 
spectrum. The electric charges and velocities of these fission fragments are 
$Q_1 e, Q_2 e$ and $V_1, V_2$, respectively. To simplify all formulas below, we 
shall assume that the both fission fragments move along the $Z-$axis. 
Furthermore, in this study we restrict ourselves to the consideration of the
non-relativistic processes only (i.e. the two velocities $V_1, V_2$ are significantly 
less than the speed of light $c$ in vacuum).
The case of arbitrary velocities is discussed in \cite{Fro07}.

The electron acceleration is directly related to the second time-derivative of the 
dipole moment, i.e.  $\ddot{\bf d} = e \ddot{\bf r}$, where ${\bf d}$ is is the dipole
moment. In the case of fission-type reaction in atomic systems the explicit formula for 
electron's acceleration is
\begin{equation}
 \ddot{\bf r} = \frac{1}{m_e} \Bigl[ \nabla \Bigl( \frac{Q_1 e^2}{R_1}
 \Bigr) + \nabla \Bigl( \frac{Q_2 e^2}{R_2} \Bigr) \Bigr] =
 \frac{Q_1 e^2}{m_e} \nabla \Bigl( \frac{1}{R_1} \Bigr) +
 \frac{Q_2 e^2}{m_e} \nabla \Bigl( \frac{1}{R_2} \Bigr) \label{dipole}
\end{equation}
where $R_1 = \sqrt{x^2 + y^2 + (z - V_1 t)^2}$ and $R_2 = \sqrt{x^2 + y^2 +
(z + V_2 t)^2}$. The notations $V_1$ and $V_2$ stand for the velocities of
the two fission fragments. The second time-derivative of the dipole moment
is
\begin{equation}
 \ddot{\bf d} = \frac{Q_1 e^3}{m_e} \nabla \Bigl( \frac{1}{R_1} \Bigr) +
 \frac{Q_2 e^3}{m_e} \nabla \Bigl( \frac{1}{R_2} \Bigr) = -
 \frac{Q_1 e^3}{m_e} \frac{{\bf R_1}}{R^{3}_1} - \frac{Q_2 e^3}{m_e}
 \frac{{\bf R_2}}{R^{3}_2}
\end{equation}
where ${\bf R}_1 = (x, y, z - V_1 t)$ and ${\bf R}_2 = (x, y, z + V_2 t)$
are the three-dimensional vectors. The intensity of the non-relativistic
bremsstrahlung from a fission-type reaction in a one-electron atomic system is
\begin{equation}
 dI = \frac{1}{4 \pi c^3} (\ddot{\bf d} \times {\bf n})^2 d\Omega =
 \Bigl( \frac{e^2}{m_e} \Bigr)^2 \frac{Q^2 e^2}{4 \pi c^3} \Bigl[
 \frac{Q_1}{Q} \frac{{\bf R_1}}{R^{3}_1} + \frac{Q_2}{Q} \frac{{\bf
 R_2}}{R^{3}_2} \Bigr]^2 sin^2\theta d\Omega \label{dI}
\end{equation}
where $\theta$ is the angle between the vector $\ddot{\bf d}$ and vector
${\bf n}$ which designates the direction of propagation of radiation. The
notation $Q$ in the last equation stands for an arbitrary electric charge.
This value can be considered as an output parameter. In particular, one can
choose $Q = Q_1$, or $Q = Q_2$. For a $N_e-$electron atomic system the last
formula takes the form
\begin{equation}
 dI = \Bigl( \frac{e^2}{m_e} \Bigr)^2 \frac{N_e Q^2 e^2}{4 \pi c^3} \Bigl[
 \frac{Q_1}{Q} \frac{{\bf R_1}}{R^{3}_1} + \frac{Q_2}{Q} \frac{{\bf
 R_2}}{R^{3}_2} \Bigr]^2 sin^2\theta d\Omega
\end{equation}
This formula can be re-written in the form
\begin{equation}
 \frac{dI}{d\Omega} = \Bigl( \frac{e^2}{m_e} \Bigr)^2 \frac{N_e Q^2 e^2}{4
 \pi c^3} \Bigl[ \frac{Q_1^2}{Q^2} \frac{1}{R^{4}_1} + \frac{Q^2_2}{Q^2}
 \frac{1}{R^{4}_2} + \frac{Q_1 Q_2}{Q^2} \Bigl(\frac{{\bf R_1} \cdot {\bf
 R_2}}{R^{3}_1 R^{3}_2}\Bigr) \Bigr]^2 sin^2\theta \label{diff}
\end{equation}
for the differential cross-section. Our derivation of these formulas is
based on the fact that radiation emitted by different post-atomic electrons
is non-coherent. Indeed, the three free electrons arising during the reaction, 
Eq.(\ref{e1}), are independent quantum systems and they move as `random' 
particles. It can be shown that the `randomization' of the phases of the
electron wave functions is directly related to the electron-electron 
repulsion which was ignored in Eqs.({\ref{dipole}}) - (\ref{dI}). At certain
conditions bremsstrahlung from fission-type process in atomic systems can be 
coherent. For instance, if two fission fragments move with very large 
velocities, while all post-atomic electrons almost do not move at the 
beginning of the process. In this case, one needs to introduce an additional 
factor $N_e$ in the last formula.

Let us discuss the spectrum of the emitted radiation. As follows from the
formula, Eq.(\ref{diff}), the intensity of bremsstrahlung from fission-type
processes in atomic systems rapidly decreases with the time $I \simeq
t^{-4}$ after the nuclear reaction ($t = 0$). Such a behavior is typical
for all finite-time process/reactions with accelerated electrons. The spectral 
resolution $R(\omega)$ (or spectral function) of the intensity of dipole 
radiation is written in the form
\begin{equation}
 dR(\omega) = \frac{4}{3 c^3} \mid \ddot{{\bf d}}_{\omega} \mid^2
 \frac{d\omega}{2 \pi} = \frac{4 \omega^4}{3 c^3} \mid {\bf d}_{\omega}
 \mid^2 \frac{d\omega}{2 \pi} \label{spectr}
\end{equation}
where ${\bf d}_{\omega}$ is the Fourier component of the dipole moment ${\bf d}$ 
defined above. As follows from Eq.(\ref{spectr}) one needs to 
find the Fourier components of the vector of the dipole moment $\ddot{{\bf 
d}}_{\omega}$, Eq.(\ref{dipole}). Finally, the problem is reduced to the calculation 
of the two following Fourier transformations
\begin{equation}
 I_1(\omega; \frac{a}{V},V,\cos\eta) = \int_0^{+\infty}
 \Bigl[\frac{a^2}{V^2} \pm 2 (\frac{a}{V} \cos\eta) \cdot t +
 t^2 \Bigr]^{-\frac32} \exp(\imath \omega t) dt \label{ft1}
\end{equation}
and
\begin{equation}
 I_2(\omega; \frac{a}{V},V,\cos\eta) = \int_0^{+\infty}
 \Bigl[\frac{a^2}{V^2} \pm 2 (\frac{a}{V} \cos\eta) \cdot t +
 t^2 \Bigr]^{-\frac32} t \exp(\imath \omega t) dt \label{ft2}
\end{equation}
where $a$ is the `effective' radius of the original electron shell and $\eta$ is the 
angle between the electron acceleration and $Z-$axis (or the line along which the two 
nuclei are moving). Note that the spectrum of the emitted radiation depends upon 
$\cos\eta$, as for all fission-type reactions and processes. The lower 
limit in these formulas is zero, but, in reality, we cannot use times $t$ which are 
shorter than $\tau = \frac{a}{c}$. Formally, this means that the lower limits in
Eqs.(\ref{ft1}) and (\ref{ft2}) can slightly be changed and this can be used
to simplify the explicit formulas for the corresponding Fourier components. The actual 
upper limits in Eqs.(\ref{ft1}) and (\ref{ft2}) are also finite. For instance, it is 
possible to obtain a very good approximation for the integrals, Eqs.(\ref{ft1}) and 
(\ref{ft2}), by using the upper limit $\frac{10 a}{V}$. In actual situations one can use 
the formula $\exp(\imath x) = \cos x + \imath \sin x$ and then calculate all arising 
integrals numerically, or by using analytical formulas from the Tables of Fourier
$sine/cosine$ transformations (see, e.g., \cite{Transf}). 

To produce the explicit formula for the spectral functions $I_1(\omega; 
\frac{a}{V},V,\cos\eta)$ and $I_2(\omega; \frac{a}{V},V,\cos\eta)$ we note 
the following relation between these two functions:
\begin{equation}
 I_2(\omega; \frac{a}{V},V,\cos\eta) = -\imath \frac{\partial}{\partial 
 \omega} I_1(\omega; \frac{a}{V},V,\cos\eta) \label{ft3}
\end{equation}
This formula substantially simplifies the computation of the Fourier transformations. 
Moreover, a very good approximation for the $I_1(\omega; \frac{a}{V},V,\cos\eta)$ 
function is:
\begin{equation}
 I_1(\omega; \frac{a}{V},V,\cos\eta) = \exp\Bigl( -\imath \omega \frac{a}{V},
 \cos\eta\Bigr) \Bigl[ \frac{\omega}{b} K_1(b \omega) \label{ft4} \\ 
 + \frac{1}{\omega} \ln \Bigl(\frac{1 + \frac12 b \omega}{1 - \frac12 b 
 \omega}\bigr) - \frac{\pi}{4} \frac{\omega^2}{(1 - \frac12 b \omega)^2}
 \Bigr] \nonumber
\end{equation}
where $b = \frac{a \sin\eta}{V}$ and $K_1(x)$ is the MacDonald function (see,
e.g., \cite{Watson}). The formula, Eq.(\ref{ft4}), describes the spectrum of 
radiation emitted by one electron accelerated by rapidly moving atomic 
fragments from the nuclear $(n,{}^6{\rm Li};{}^4{\rm He},t)-$reaction. The actual 
spectra emitted during the nuclear $(n,{}^6{\rm Li};{}^4{\rm He},t)-$reaction in the
three-electron Li atom can be different from the spectrum predicted by the formula 
Eq.(\ref{ft4}) since we have ignored all contributions from the electron-electron
repulsion(s). It is interesting to study such radiation in detail. In general, the 
intensity of bremsstrahlung from the reaction, Eq.(\ref{e1}), can be amplified by
accelerating the incident ${}^6$Li atom or ${}^6$Li$^{+}$ ion to very large velocities. 

\section{Conclusion}

We have considered the nuclear reaction, Eq.(\ref{e1}), with the thermal/slow neutrons 
in the three-electron Li atom. It is shown that a number of different atoms and ions
can be formed during this reaction. This also includes the formation of the electrically 
charged tri-electron $e^{-}_3$ ion which is a quasi-stable negatively charged ion.
It has the same (doublet) symmetry as the wave function of the incident Li atom. We 
have developed an effective method for accurate evaluation of the final state 
probabilities, i.e. probabilities to form  different atomic species in a variety of 
bound states. This method is based on the use of special three- and two-electron 
wave functions of the Li and He atoms, respectively, whose explicit form allows one to 
obtain accurate numerical results for all bound state 
properties of the incident and final atoms/ions. On the other hand, for the same wave
functions it is easy to perform all complete and/or partial Fourier transformations 
needed in calculations of the final state probabilities. The spectrum of the emitted
radiation has been investigated. It is shown that such a spectrum essentially coincides 
with the known spectrum of bremsstrahlung emitted by the accelerated electrons. 

Our procedure can now be used for the more complicated nuclear $(n;\alpha)-$reaction in 
the five-electron B-atom (see Appendix). The $(n,{}^{10}{\rm B};{}^{7}{\rm Li},
\alpha)-$reaction on this atom is of great interest in a number of applications, e.g., 
in medical physics (BNCT). The Li-atom/ion and He-atom/ion which are formed during this 
reaction may contain up to three and two electrons, respectively. These atomic fragments 
move rapidly, with the velocities $v_{Li} \approx 2.40896$ $cm \cdot sec^{-1}$ and 
$v_{\alpha} \approx 4.21568$ $cm \cdot sec^{-1}$ in the case of slow neutrons. The sudden 
approximation can certainly be applied to the He-atom and He-like ions. However, this 
approximation cannot be used for internal electrons (or $1^2s-$electrons) of the Li 
atom/ion, since the velocities of these two electrons are comparable with the final 
velocity of the ${}^{7}$Li nucleus. It is very likely that the probability to observe the 
${}^{7}$Li$^{2+}$ ions after Eq.(\ref{e2}) will be relatively small $\le$ 5 - 10 \%, while
analogous probability for the ${}^{7}$Li$^{+}$ ions will be relatively large ($\ge$ 50 \%). 

\begin{center}
    {\bf Appendix}
\end{center}

In this study our analysis was restricted to the consideration of the nuclear 
$(n;t)-$reaction, Eq.(\ref{e1}), in the three-electron Li atom. Moreover, such a 
reaction is considered for thermal and slow neutrons only. In reality, the 
nuclear reaction of the ${}^{6}$Li nucleus, Eq.(\ref{e1}), proceeds with 
neutrons of all energies and the energy released increases almost linearly 
with the energy of the incident neutron. For fast neutrons with $E_n \ge 1$ 
$MeV$ the energy released is substantially different from the value quoted
in Eq.(\ref{e1}) and the velocities of the atomic fragments increase
correspondingly. Investigation of the final state (atomic) probabilities and
bremsstrahlung emitted during such processes with fast neutrons is of interest 
in some applications. Note that the cross-section of this reaction has a large
resonance (maximum) $\sigma \approx 4.5$ $barn$ at $E_n \approx 240 - 270$
$keV$, but it is relatively large for neutrons of all energies $E_n \le 0.8$
$MeV$. This makes the reaction, Eq.(\ref{e1}), extremely important in the
thermonuclear ignition and following propagation of the thermonuclear
burning wave in highly compressed ($\rho \ge 100$ $g \cdot cm^{-3}$)
${}^6$LiD deuteride which is routinely used as a fuel in
modern thermonuclear explosive devices (see, e.g., \cite{Gon},
\cite{Fro98}). The analogous ($n,{}^{3}$He;$p,t$)-reaction \cite{FrWa09} also
plays a very important role in such processes.

In general, the nuclear $(n,t)$-reactions of the ${}^{6}$Li and ${}^{3}$He
nuclei with neutrons allow one to reduce drastically the overall bremsstrahlung 
loss from the hot combustion zone and increase the tritium/deuterium ratio which 
is crucially important to start new thermonuclear $(d,t)-$reactions. Briefly, 
by using the ${}^{6}$LiD deuteride in modern thermonuclear explosive devices we 
can reduce the required compressions to relatively small values. In many cases 
such compressions are dozens times smaller (usually, in 25 - 40 times smaller 
\cite{Fro98}) than compressions required for any other (solid) thermonuclear 
fuel, e.g., ${}^{7}$LiD deuteride. On the other hand, by using compressions which 
are provided by a standard `primary' nuclear charge one can create extremely
compact thermonuclear explosive devices based on ${}^{6}$LiD deuteride.
The idea to use pure ${}^{6}$LiD deuteride in thermonuclear explosive
devices was originally proposed by V.L. Ginzburg in 1948-1949 (see
discussion and references in \cite{Gon}).

Some other $(n;t)-, (n;p)-$ and $(n;\alpha)-$reactions are often used in different 
applications. For instance, the nuclear reaction of ${}^{10}$B nuclei with 
slow neutrons
\begin{equation}
 {}^{10}{\rm B} + n = {}^7{\rm Li} + {}^4{\rm He} + 2.791 \; \; MeV
 \label{e2}
\end{equation}
is extensively used in the boron neutron capture therapy (BNCT, for short), 
or boron neutron-capture synovectomy \cite{BNCT1} - \cite{Haw3}, to treat 
different forms of cancer, including brain cancer. The fast $\alpha-$particle 
produced in the reaction, Eq.(\ref{e2}), kills (or at least `badly damages') 
one cancer cell before it finally stops. The modern applications of this 
reaction to cancer treatment are based on the use of molecules which contain a 
large number of ${}^{10}$B-atoms, e.g., the Na$_3$[B$_{20}$ H$_{17}$NH$_{2}$
CH$_{2}$CH$_{2}$ NH$_{2}$] molecule, other similar molecules, and molecular
clusters \cite{Haw1}, \cite{Haw2} (see also \cite{Haw3} and references
therein). In this case the overall energy release from the reaction,
Eq.(\ref{e2}), in one cancer cell can be extremely large. Correspondingly,
the local temperature in the whole cell suddenly increases to very large
values and this kills the incident cancer cell with almost 100 \%
probability. Note that the tritium nucleus does not form in the nuclear 
reaction Eq.(\ref{e2}) which means that it is safe to initiate this reaction 
inside of a human body. By studying the nuclear reaction, Eq.(\ref{e1}), in
few-electron atoms and ions we want to develop a number of reliable
theoretical methods and numerical procedures which can be later used in
applications to the analogous reaction, Eq.(\ref{e2}).

\newpage
  \begin{table}[tbp]
   \caption{An example of the trial, three-electron wave function
            constructed with the use of $N = 23$ semi-exponential radial
            basis functions, Eq.(7). This wave function produces the total
            energy $E$ = -7.44859276608 $a.u.$ for the ground ${}^2S-$state
            of the ${}^{\infty}$Li atom. Only one electron spin-function
            $\chi_1 = \alpha \beta \alpha - \beta \alpha \alpha$ was used
            in these calculations.}
     \begin{center}
     \scalebox{0.72}{%
     \begin{tabular}{cccccccc}
      \hline\hline
 $k$ & $m_1(k)$ & $m_2(k)$ & $m_3(k)$ & $C_{k}$ & $\alpha_k$ & $\beta_k$ & $\gamma_k$ \\
     \hline
 1  & 0 & 0 & 1 &   0.146131429481911E+02 & 0.416423958308045E+01 & 0.390330393520923E+01 & 0.754647967615054E+00 \\
 2  & 1 & 0 & 1 &   0.110681883475133E+03 & 0.526120133598745E+01 & 0.699723209902291E+01 & 0.604874234309236E+00 \\
 3  & 1 & 1 & 1 &   0.234004890152000E+03 & 0.598950989816454E+01 & 0.378246213278597E+01 & 0.590931685390717E+00 \\
 4  & 2 & 0 & 1 &   0.116882861303726E+03 & 0.413820964104499E+01 & 0.322650638788059E+01 & 0.590283203974117E+00 \\
 5  & 0 & 0 & 0 &  -0.112533836667930E+02 & 0.404234943769060E+01 & 0.402575246908481E+01 & 0.136870242306376E+01 \\
 6  & 0 & 0 & 2 &   0.124148253762594E+01 & 0.772078877480103E+01 & 0.208089611437233E+01 & 0.906692873308709E+00 \\
 7  & 0 & 0 & 3 &   0.129950444032051E+01 & 0.125634586976915E+02 & 0.167180833975207E+02 & 0.111484879880926E+01 \\
 8  & 3 & 0 & 1 &   0.121940082563610E+02 & 0.336154087416427E+01 & 0.329089537444583E+01 & 0.569461303170668E+00 \\
 9  & 2 & 2 & 1 &  -0.815264312642982E+01 & 0.315618697991587E+01 & 0.293590042487235E+01 & 0.642752252837086E+00 \\
 10 & 0 & 0 & 4 &   0.231307852206996E-01 & 0.216119418842614E+01 & 0.354528921898964E+01 & 0.848100637437686E+00 \\
 11 & 1 & 0 & 0 &  -0.290323204459308E+02 & 0.303111504960694E+01 & 0.340407714984374E+01 & 0.439831339492880E+00 \\
 12 & 1 & 0 & 2 &  -0.782805087195444E+01 & 0.319874342329211E+01 & 0.294880236447181E+01 & 0.670201325893484E+00 \\
 13 & 4 & 0 & 1 &   0.466451849833029E+05 & 0.257191421047538E+02 & 0.198922117452300E+02 & 0.745721864825035E+00 \\
 14 & 5 & 0 & 1 &   0.102312737866967E+03 & 0.610735801244370E+01 & 0.126738573321754E+02 & 0.583192334409026E+00 \\
 15 & 1 & 1 & 0 &   0.496831723313492E+01 & 0.214720891197913E+01 & 0.139151014500059E+02 & 0.332004403600092E+00 \\
 16 & 1 & 1 & 2 &   0.331570408306929E+01 & 0.313893628304852E+01 & 0.288670468153774E+01 & 0.656255394173358E+00 \\
 17 & 2 & 0 & 0 &  -0.224488896165048E+03 & 0.111494404746535E+02 & 0.712253949382151E+01 & 0.412242317869619E+00 \\
 18 & 2 & 0 & 2 &  -0.200426722868461E+01 & 0.282336074847368E+01 & 0.292360100848728E+01 & 0.638644442755579E+00 \\
 19 & 0 & 0 & 5 &   0.186252350189224E-02 & 0.103900277658347E+02 & 0.241603283627644E+01 & 0.858660133717424E+00 \\
 20 & 2 & 1 & 1 &   0.314011900712871E+01 & 0.314575640023816E+01 & 0.238785024725699E+01 & 0.645792868876097E+00 \\
 21 & 3 & 0 & 2 &   0.164525876141841E+01 & 0.318702139403640E+01 & 0.348200049008609E+01 & 0.663487519449685E+00 \\
 22 & 3 & 0 & 0 &   0.549306690646211E+01 & 0.313423005590041E+01 & 0.376361954006700E+01 & 0.494738425712773E+00 \\
 23 & 3 & 1 & 1 &   0.489167893642778E+01 & 0.310583281642480E+01 & 0.299668864388672E+01 & 0.617551331047668E+00 \\
  \hline\hline
  \end{tabular}}
  \end{center}
  \end{table}
  \begin{table}[tbp]
   \caption{The probabilities (in \%) of observing selected electron 
            final states in the helium ion ${}^4$He$^{+}$ and tritium 
            atom ${}^3$H arising in the exothermic nuclear reaction 
            Eq.(1) of the three-electron lithium-6 atom in its ground 
            ${}^2S-$state with slow neutrons.}
     \begin{center}
     \begin{tabular}{llll}
        \hline\hline
 atom/state & $1s$   &  $2s$    &  $2p$  \\
           \hline
 ${}^{4}$He$^{+}$ & 9.52386$\cdot 10^{-2}$ & 1.66605$\cdot 10^{-2}$  &  6.60170$\cdot 10^{-4}$ \\

 ${}^{3}$H & 2.49810$\cdot 10^{-3}$ & 3.05880$\cdot 10^{-4}$  &  2.00251$\cdot 10^{-6}$ \\
        \hline\hline
  \end{tabular}
  \end{center}
  \end{table}

\end{document}